\documentclass[english,aps,prl,twocolumn,amsmath,amssymb,showpacs,superscriptaddress,notitlepage]{revtex4-2}

\usepackage[T1]{fontenc}
\usepackage{color}
\definecolor{page_backgroundcolor}{rgb}{1, 1, 1}
\pagecolor{page_backgroundcolor}
\usepackage{amsmath}
\usepackage{graphicx}
\usepackage{microtype}
\usepackage{bm}
\usepackage{makecell}
\usepackage[normalem]{ulem}
\usepackage{enumitem}
\usepackage[marginal]{footmisc}
\usepackage{braket}
\usepackage{pifont}
\usepackage{float}

\makeatletter
\usepackage{amssymb}
\usepackage{amsmath}
\usepackage{graphicx}
\usepackage[colorlinks=true,linkcolor=blue,anchorcolor=red,citecolor=blue,urlcolor=blue]{hyperref}
\usepackage[caption=false]{subfig}
\usepackage{lipsum}

\makeatother

\usepackage{babel}

\begin{document}
\global\long\def\figurename{Fig.}%

\title{Floating Edge Bands in the Bernevig-Hughes-Zhang model with Altermagnetism}

\author{Yang-Yang Li}
\affiliation{Department of Modern Physics, University of Science and Technology of China, Hefei, Anhui 230026, China}
\affiliation{International Center for Quantum Design of Functional Materials (ICQD),
University of Science and Technology of China, Hefei, Anhui 230026, China}

\author{Song-Bo Zhang}
\email{songbozhang@ustc.edu.cn}
\affiliation{Hefei National Laboratory, University of Science and Technology of China, Hefei 230088, China}
\affiliation{International Center for Quantum Design of Functional Materials (ICQD),
University of Science and Technology of China, Hefei, Anhui 230026, China}

\begin{abstract}
Floating edge bands (FEB) have been identified in systems such as obstructed atomic insulators and layered nonsymmorphic semimetals, attracting considerable interest recently. 
Here we demonstrate that FEB can arise in a simplified model incorporating altermagnetism. By enhancing the Bernevig-Hughes-Zhang model on a square lattice with additional altermagnetic and Zeeman fields perpendicular to the 2D plane, we uncover the emergence of FEB that are distinct from the bulk bands across the entire Brillouin zone and over broad parameter regimes. 
We calculate topological phase diagrams, highlighting the strong topological properties characterized by the Chern number and the weak topological properties marked by the winding number. Furthermore, we provide analytical results of the energy spectrum and the wave functions of the FEB. We also study the robustness of the FEB, showcasing its resilience against various perturbations such as geometric rotation, energy spectrum asymmetry, and spin coupling. Our findings advance our understanding of FEB and may pave new avenues for further exploration of topological phases in quantum materials.
\end{abstract}

\maketitle

\section{Introduction}
Conventionally, edge bands bridge  (and merge into) the bulk conduction and valence bands, and exist only in a limited range of momentum space~\cite{Hasan10RMP,XLQi11RMP,Shen2012topological}. Recently, it has been found that edge bands can emerge over the entire Brillouin zone and may detach from the bulk bands, floating inside the band gap~\cite{PhysRevX.7.041073,Zhu2018,Liu2021,2019arXiv191106799L, 2020arXiv201215220C,https://doi.org/10.1002/adma.202101591,PhysRevB.105.165135,PhysRevB.108.085135,Liu2023,PhysRevB.107.174101,Nuñez_2024,PhysRevB.106.155144,PhysRevB.108.L100101,CALi22PRB,PhysRevB.108.L100101,https://doi.org/10.1002/adfm.202316079,2024arXiv240605380L}. This exotic band structure has been identified in various systems, including layered nonsymmorphic semimetals \cite{PhysRevX.7.041073,Zhu2018,Liu2021,2019arXiv191106799L, 2020arXiv201215220C,https://doi.org/10.1002/adma.202101591} and obstructed atomic insulators \cite{PhysRevB.105.165135,PhysRevB.108.085135,Liu2023,PhysRevB.107.174101,Nuñez_2024,PhysRevB.106.155144,PhysRevB.108.L100101,CALi22PRB,PhysRevB.108.L100101,https://doi.org/10.1002/adfm.202316079,2024arXiv240605380L}, and attracted broad attention. With their unique properties, floating edge bands (FEB) are robust against back scattering and offer a platform for frequency multiplexing transport~\cite{Ma:24}. They are also closely related to other intriguing topological phases, such as higher-order topological states~\cite{PhysRevB.109.245306,CALi22PRB,LiCA23PRB,PhysRevB.106.155144,PhysRevB.108.L100101,PhysRevB.107.174101,PhysRevB.108.L100101,Nuñez_2024}.

The Bernevig-Hughes-Zhang (BHZ) model is a seminal model for understanding topological phases in two dimensions (2D). 
Initially introduced to capture the quantum spin Hall effect~\cite{doi:10.1126/science.1133734,CXLiu08PRL,doi:10.1126/science.1148047,PhysRevLett.101.246807,PhysRevLett.107.136603} and quantum anomalous Hall effect \cite{PhysRevLett.101.146802} in systems like HgTe/CdTe quantum wells, it has since evolved into a generalized model for examining various topological phenomena. The model is characterized by a mass term $M-Bk^2$ in its low-energy approximation. In the original form, the sign of $MB$ distinguishes the topological phase ($MB>0$) from the trivial phase ($MB<0$). With additional terms, the model has been applied to describe many other topological phases, such as topological Anderson insulators \cite{JLi09PRL,Groth09PRL,HJiang09PRL,CZChen15PRB}, Floquet topological insulators \cite{PhysRevB.101.174314,Zhu_2014,PhysRevB.96.054207}, and flat bands \cite{PhysRevB.82.085118,Imura2011,PhysRevB.104.L041106}, further broadening its applicability and deepening our understanding of topological phases of matter.

Recently, a third type of collinear magnetic order called altermagnetism has been observed in various materials~\cite{2024arXiv240602123B,naka2019spin,doi:10.7566/JPSJ.88.123702,PhysRevB.102.144441, PhysRevB.102.014422,PhysRevMaterials.5.014409,doi:10.1073/pnas.2108924118, PhysRevX.12.021016,PhysRevX.12.031042,PhysRevX.12.040501,PhysRevB.107.L100418,Hariki_2024,krempasky2024altermagnetic,Ma2021}. This novel magnetic phase may be caused by Fermi liquid instabilities~\cite{CJWu07PRB,Soto-Garrido14PRB,Ahn19PRB}. It can also emerge from crystal symmetries that alternate its spin polarization in both coordinate and momentum space while maintaining net zero magnetization~\cite{PhysRevX.12.031042,PhysRevX.12.040501}. Unlike conventional ferromagnetism and antiferromagnetism, altermagnetism possesses non-relativistic spin splitting which anisotropically depends on momentum, e.g., analogous to $d$-wave pairing potential in superconductors~\cite{PhysRevX.12.040501}. These distinctive properties prompt several exotic phenomena~\cite{PhysRevB.109.245306,DZhu23PRB, PhysRevLett.131.076003,Zhang2024,Beenakker23PRB,HPSun24arXiv,Nagae2024arXiv,PhysRevB.108.054511,PhysRevB.108.L060508,2023arXiv230609413G,Zhou_2024,YXLi23PRB,MMWei24PRB,QCheng24PRB,Sumita23PRB,Zyuzin24PRB,2024arXiv240418616H,Banerjee24PRB,2024arXiv240510656L,2024arXiv240808297D,2024arXiv240804459A,2024arXiv240810413S}, such as anomalous Hall effects~\cite{doi:10.1126/sciadv.aaz8809,Feng2022,PhysRevLett.130.036702,Reichlova2024}, giant tunneling magnetoresistance~\cite{DFShao21NC,PhysRevX.12.011028}, 
and zero-field finite-momentum Cooper pairing \cite{Zhang2024,Sumita23PRB,2023arXiv230914427C,2024arXiv240702059H,Sim2024arXiv}.

In this work, we investigate the influence of $d$-wave perpendicular altermagnetic and Zeeman fields on the BHZ model on a 2D square lattice. 
We identify broad parameter regimes in which FEB can emerge and detach from the bulk bands throughout the Brillouin zone. We obtain the phase diagrams of the altermagnetic BHZ model, using Chern numbers and winding numbers at high-symmetry points [see Fig.~\ref{fig:PhaseDiagram}].  
If the system is in a trivial state at zero fields, altermagnetism alone cannot generate FEB phase. However, the presence of a Zeeman field guarantees a FEB phase by tuning the altermagnetism, regardless of the original state of the BHZ model. 
Moreover, the FEB generated solely by altermagnetism is tangent to the bulk bands, with either additional low-energy modes with different spin polarization [see Figs.~\ref{fig:DOS}(b) and \ref{fig:DOS}(c)] or some spin degeneracy over a certain range of momenta along the edge [see Figs.~\ref{fig:DOS}(c) and \ref{fig:DOS}(d)]. With a Zeeman term, the bulk bands can invert and completely detach from the FEB, forming a suspending edge band (SEB). For the lattice model in a ribbon geometry, we analytically derive the energy spectrum and wave functions of the floating edge states. We find that the localization length of the floating edge states is essentially independent of the momentum along the edge, in stark contrast to the edge states in quantum spin Hall insulators.
Finally, we show that the FEB phase is rather robust against perturbations such as a rotation of the ribbon orientation, spectrum asymmetries, and spin coupling (e.g., by an in-plane Zeeman field or superconductivity) in the system.

The remainder of the article is organized as follows. In Sec.~\ref{sec:AlterBHZ}, we introduce the altermagnetic BHZ model. In Sec.~\ref{sec:FEB}, we calculate the phase diagram, analyze the diverse topological phases in the system, and derive the analytical energy spectrum and wave function of the edge states. In Sec.~\ref{sec:robustness}, we demonstrate the robustness of the FEB phase against various perturbations. Finally, we conclude our results in Sec.~\ref{sec:summary}.

\section{BHZ model with $d$-wave altermagnetism \label{sec:AlterBHZ}}

We consider the BHZ model on a 2D square lattice~\cite{PhysRevLett.101.146802}. Under the presence of $d$-wave altermagnetic and Zeeman fields applied perpendicular to the lattice plane, the model Hamiltonian in momentum space reads
\begin{equation}
\begin{aligned}
  H({\bf k})= & \; M({\bf k})\sigma_3+A(\sin k_x s_3\sigma_1 + \sin k_y \sigma_2)\\
  & +2J(\cos k_y -\cos k_x)s_3\sigma_3 + G s_3 \sigma_3,
  \label{eq:H}
\end{aligned}
\end{equation}
where $M({\bf k})=M_0+2B(\cos k_x + \cos k_y)$ with $M_0=M-4B$ and ${\bf k}=(k_x,k_y)$ being the Bloch wave vector in the plane. It is written in the basis $\Psi=(c_{\uparrow,a},c_{\uparrow,b},c_{\downarrow,a},c_{\downarrow,b})^T$ with subscripts $a$ and $b$ distinguishing two orbitals of opposite parity. $\bm{s}=(s_1,s_2,s_3)$ are Pauli matrices for spins and 
 $\bm{\sigma}=(\sigma_1,\sigma_2,\sigma_3)$ are Pauli matrices for the two orbitals. The first line in Eq.~\eqref{eq:H} is the original BHZ model, where the $A$ term accounts for orbital-momentum locking and the $M({\bf k})$ term is the mass term. The $J$ and $G$ terms describe the $d$-wave altermagnetic and Zeeman fields, respectively. While the Zeeman field $G$ can be introduced by applying external magnetic fields, the altermagnetic field can be achieved via proximity effect by coupling the BHZ system to an altermagnetic material. Alternatively, the altermagnetism can be realized and controlled using antiferroelectricity~\cite{2024arXiv241006071D}. Without loss of generality, we consider $A>0$~\footnote{For a negative $A$, the topological invariants flip their signs.}. We set the lattice constant to be unity. To illustrate the key results, we further assume that the two orbitals have opposite responses to the fields. However, we note that the main results extend beyond this unessential assumption, as we will show later [Sec.~\ref{sec:robustness}]. In the BHZ model, the two spin states are uncoupled. Thus, the Hamiltonian split into two $2\times2$ blocks even in the presence of the altermagnetic and Zeeman fields in the $z$ direction. The two blocks can be related by 
\begin{equation}
    H_-({\bf k},J,G)=H_+^*(-{\bf k},-J,-G).\label{eq:+-}
\end{equation}
In the absence of altermagnetism ($J=0$), each block of the Hamiltonian resembles the Qi-Wu-Zhang model~\cite{Qi-Wu-Zhang06PRB}. This block-diagonal property in spin space allows us to deal with two spins separately. 
The bulk energy spectrum can be found as
\begin{subequations}
\begin{align}
E_{1(2)}=\pm \sqrt{A^2(\sin^2 k_x+\sin^2 k_y)+\epsilon^2_{J+}},\\
E_{3(4)}=\pm \sqrt{A^2(\sin^2 k_x+\sin^2 k_y)+\epsilon^2_{J-}},
\end{align}
\end{subequations}
where $\epsilon_{J\pm}=M_0 \pm G +2(B\pm J)\cos k_x+2(B\mp J)\cos k_y$.

The bare BHZ model ($J=G=0$) possesses time-reversal symmetry and four-fold rotation symmetry, as indicated by $\mathcal{T}H({\bf k})\mathcal{T}^{-1}=H(-{\bf k})$ and $\mathcal{C}_{4z}H(k_x,k_y)\mathcal{C}_{4z}^{-1}=H(k_y,-k_x)$, respectively. Here, $\mathcal{T}=-i s_2\mathcal{K}$ with complex conjugation $\mathcal{K}$ is the time-reversal operator, and  $\mathcal{C}_{4z}=e^{i\frac{\pi}{4}\sigma_3}$ is the fourfold rotational symmetry about the $z$ axis. 
When $|M_0|<4|B|$, the model is in a topological phase, exhibiting a pair of $\mathcal{T}$-protected helical edge states. Otherwise, it is in a trivial phase.
The presence of helical edge states is dictated by a $Z_2$ invariant, as we will discuss later. 
Both the altermagnetic and Zeeman terms break time-reversal symmetry (TRS). However, when the Zeeman field is absent ($G=0$), the system preserves the combined $\mathcal{C}_{4z}\mathcal{T}$ symmetry. Thus, the two blocks are still degenerate at the $\Gamma=(0,0)$ and $M=(\pi,\pi)$ points. A finite $G$ lifts the degeneracy protected by the $\mathcal{C}_{4z}\mathcal{T}$ symmetry.

\section{Floating Edge Band \label{sec:FEB}}

\subsection{Topological Numbers}

Next, we show the emergence of FEB in the BHZ model under the altermagnetic field. We first derive the topological invariants for the model. Since the model is decoupled into two blocks, we can treat them separately. 

\begin{figure}
\includegraphics[width=0.5\textwidth]{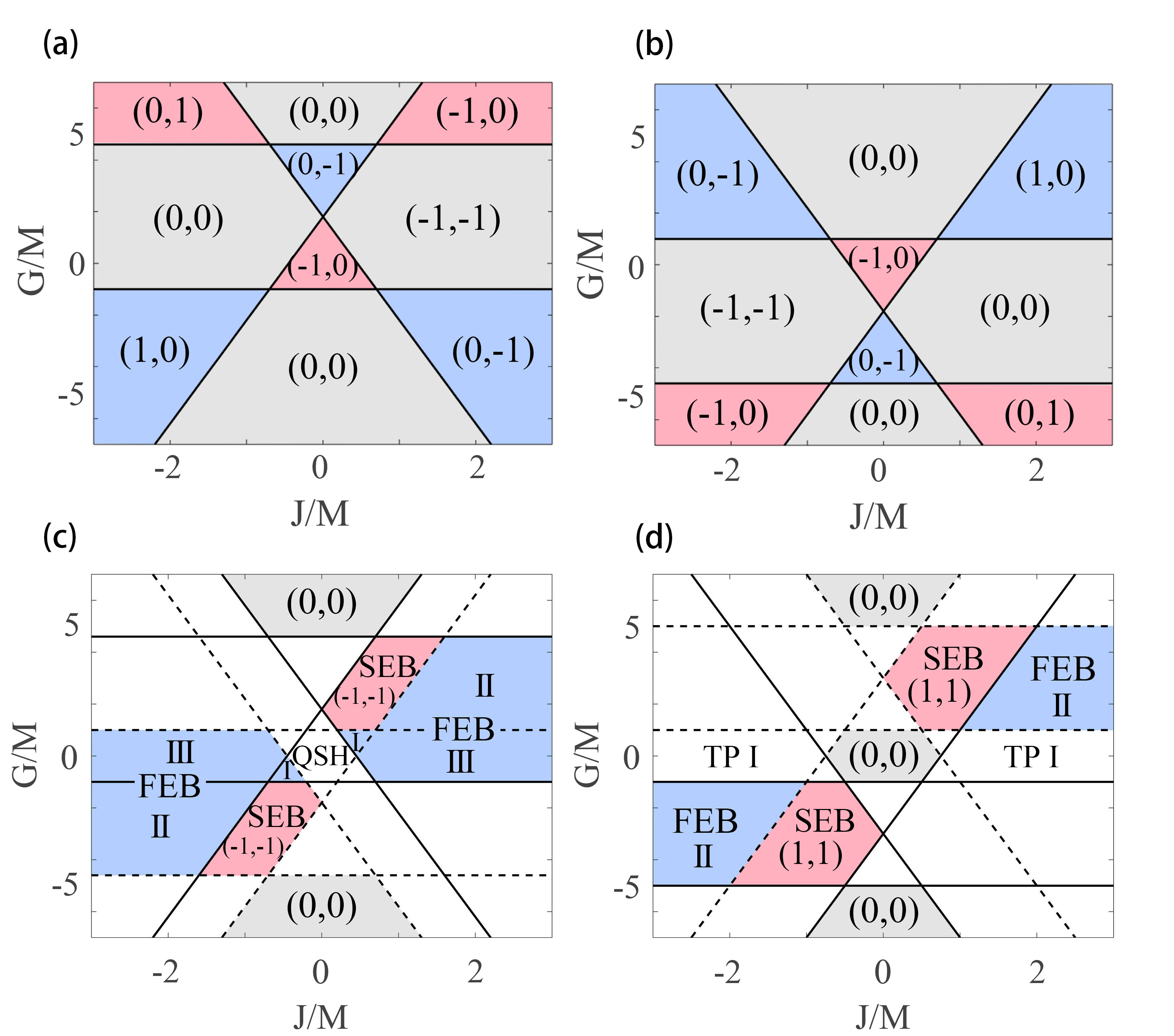}

\caption{(a) Topological phase diagram in the spin-up block classified by winding number $(W_0^{+},W_\pi^{+})$ and Chern number $C_+$ (shaded in blue/red for $C_{+}=\pm 1$, and grey for $C_{+}=0$). (b) The same as (a) but for the spin-down block. Other parameters for both (a) and (b) are $M=2$, $A=1$, and $B=1.4$. (c) Topological phase diagram of the full altermagnetic BHZ model as functions of altermagnetic $J$ and Zeeman field $G$, with parameters $M=2$, $A=1$, and $B=1.4$. The FEB phase is characterized by winding numbers $(W_0^{\pm},W_\pi^{\pm})=(-1,-1)$ for either spin ("+" or "-"). 
The SEB phase is characterized by the total winding numbers $(W_0,W_\pi)=(-1,-1)$, which is the sum over both spins. 
Solid (dashed) lines correspond to the phase transition line of the spin-up (-down) block. The FEB phase contains regions I, II, and III, corresponding to a FEB phase in one spin block while different topological phases (shown in Figs.~\ref{fig:DOS}(b-d)) in the other spin block. (d) The same as (c) but for $B=-1$. QSH and TP I denote different topological phases shown in Figs.~\ref{fig:DOS}(a) and (f), respectively.
}
\label{fig:PhaseDiagram}
\end{figure}

We take the $H_+$ block for illustration. In terms of the Pauli matrices $\bm\sigma$,  $H_+$ takes the form $H_+({\bf k})=\bm{d}({\bf k})\cdot\bm{\sigma}$, where the $\bm{d}=(d_1,d_2,d_3)$ vector is
\begin{align}
 d_1 = &\; A\sin k_x, \;\;\; d_2=A\sin k_y, \notag \\
 d_3 = & M(\bm{\mathrm{k}})+2J(\cos k_y-\cos k_x)+G.
 \end{align}
In the model, there is no TRS, particle-hole symmetry, or chiral symmetry, thus it belongs to class A~\cite{Ryu_2010} in the $\mathbb{Z}$ classification. The topological property of the two-band subblock Hamiltonian can be characterized by a Chern number defined on the first Brillouin zone~\cite{Qi-Wu-Zhang06PRB}:
\begin{equation}
    C_+=-\frac{1}{4\pi}\int_{\text{1BZ}}d^2 {\bf k} \frac{\bm{d}\cdot(\partial_{k_x}\bm{d}\times \partial_{k_y}\bm{d})}{|\bm{d}|^3},
\end{equation}
where $|\bm{d}|=\sqrt{\sum_{\nu=1}^3 d_\nu^2}$. 
A straightforward calculation yields
\begin{align}
  C_+=
   & \begin{cases}
      -1,&4|J|<G+M_0<4|B|\\
      -1,&-4|J|<G+M_0<-4|B|\\
      0,&|G+M_0|<4 \min(|B|,|J|)\\
      0,&|G+M_0|>4 \max(|B|,|J|)\\
      1,&-4|B|<G+M_0<-4|J|\\
      1,&4|B|<G+M_0<4|J|
  \end{cases}
  \label{eq:ChernNumber}
\end{align}
The Chern number for the spin-down block $H_-$ can be obtained easily by exploiting the relation in Eq.~(\ref{eq:+-}). When the magnetisms are absent (i.e., $G=J=0$), we find $C_\pm=\mp1$ for $0<M_0<4|B|$ and $C_\pm=\pm 1$ for $-4|B|<M_0<0$. These opposite Chern numbers for the two spin blocks define the $Z_2$ invariant of the BHZ model, $\mathbb{Z}_2=C_+-C_-\neq 0$, and dictate the helical edge states of the system under open boundary conditions. In other parameter regions, we have trivial Chern numbers, i.e., $C_\pm  = 0$. Accordingly, there are no edge states across the bulk gap. However, when introducing the altermagnetic and Zeeman exchange fields, topological phases emerge in more parameter regions, as indicated by Eq.~\eqref{eq:ChernNumber}. The topological phase diagram determined by the Chern numbers is shown in Figs.~\ref{fig:PhaseDiagram}(a) and \ref{fig:PhaseDiagram}(b), where the colors indicate the Chern numbers.

More topological phenomena can emerge~\cite{DZhu23PRB} when we confine the 2D system into a ribbon geometry, even when the bulk topology appears trivial based on the Chern number.
Without loss of generality, we consider a ribbon along the $x$ direction (we will consider an arbitrary angle $\theta$ rotated from the $x$ axis in Sec.~\ref{sec:robustness}).
In this geometry, the momentum $k_x$ in the $x$ direction remains a good quantum number. Therefore, we treat the Hamiltonian in Eq.~(\ref{eq:H}) as a one-dimensional Hamiltonian $H_{k_x}(k_y)$ for each $k_x$.

Although the overall chiral symmetry is not preserved in the full Hamiltonian, at high symmetry points, $k_x=0$ and $k_x=\pi$, an emergent chiral symmetry $\mathcal{C}=\sigma_1$ appears. This is illustrated by the relation $\{\mathcal{C},H_{k_x=0/\pi}\}=0$. In the generalized AZ classification~\cite{PhysRevX.14.011057,2024arXiv240709458N,2024arXiv240718273S}, the quasi-1D subsystems $H_{k_x=0/\pi}$ at these high symmetry points can be categorized as class AIII with the $\mathbb{Z}^{\text{\ding{56}}}$ classification, where Wannier localizability ("\text{\ding{56}}") enables edge states to detach from the bulk. As we will show numerically in Sec.~\ref{subs:TopPhase}, this property further facilitates the emergence of detached boundary states throughout the Brillouin zone. Thus, it is sufficient to use winding number~\cite{Ryu_2010} to classify the detached FEB phases.
To obtain the winding number, we first perform a unitary transformation $U=e^{i\frac{\pi}{4}\sigma_2}$ to diagonalize the chiral symmetry operator. After the transformation, the chiral symmetry operator becomes $\tilde{\mathcal{C}}=U\mathcal{C}U^{-1}=\sigma_3$. Accordingly, the spin-up Hamiltonian transforms to
\begin{equation}
    UH^+_{0/\pi}U^{-1}=\begin{pmatrix}
        0&q^+_{0/\pi}(k_y)\\
        [q^+_{0/\pi}(k_y)]^\dagger&0
    \end{pmatrix},
\end{equation}
where $q^{+}_{0/\pi}(k_y)= -[M_0+2B(\pm 1+\cos k_y)]-iA\sin k_y
     - 2J(\cos k_y\mp 1)- G$
with $\pm(\mp)$ corresponding to $k_x=0$ and $k_x=\pi$, respectively.
Hence, the winding numbers at $k_x=0$ and $k_x=\pi$ can be obtained as~\cite{Ryu_2010}
\begin{equation}
    W^+_{0/\pi}=\frac{i}{2\pi}\int_{-\pi}^{\pi}dk_y [q^{+}_{0/\pi}(k_y)]^{-1}\partial_{k_y}q^{+}_{0/\pi}(k_y).
\end{equation}
Explicitly, they are given by
\begin{equation}
W^+_0=\begin{cases}
    -1, &-4B<G+M_0<4J\\
    0, &G+M_0 <4\min(-B,J)\\
    0, &G+M_0 >4\max(-B,J)\\
    1, &4J<G+M_0 <-4B
    \end{cases}
\end{equation}
and
\begin{equation}
W^+_{\pi}=\begin{cases}
    -1, &-4J<G+M_0<4B\\
    0, &G+M_0<4\min(B,-J)\\
    0, &G+M_0>4\max(B,-J)\\
    1, &4B<G+M_0<-4J
    \end{cases}
\end{equation}
The winding numbers for the spin-down block $H_-$ can be obtained using the relation in Eq.~(\ref{eq:+-}). The phase diagrams for the two spin blocks, classified by Chern number and winding numbers $(W_0^\pm,W_\pi^\pm)$, are presented in Figs.~\ref{fig:PhaseDiagram}(a) and \ref{fig:PhaseDiagram}(b), respectively.  Notably, the phase boundaries defined by the Chern number and winding numbers are identical, and they can be determined by the gap closing at the $\Gamma$ and $M$ points. 

\subsection{Diverse Topological Phases \label{subs:TopPhase}}

Diverse topological phases arise in the system, as distinguished by the winding numbers $(W_0,W_\pi)$. In particular, the FEB phase emerges across broad parameter regimes, as indicated by the color-shaded areas in Figs.~\ref{fig:PhaseDiagram}(c) and \ref{fig:PhaseDiagram}(d). To illustrate the edge state characteristics, we consider the system in a ribbon geometry, with periodic boundary conditions in the $x$ direction and open boundaries in the $y$ direction. To proceed, we analyze the cases with $-4B<M_0<0$ and $M_0>-4B>0$, respectively.
For illustration, we focus on the spin-up block. 

\begin{figure}
    \includegraphics[width=1\linewidth]{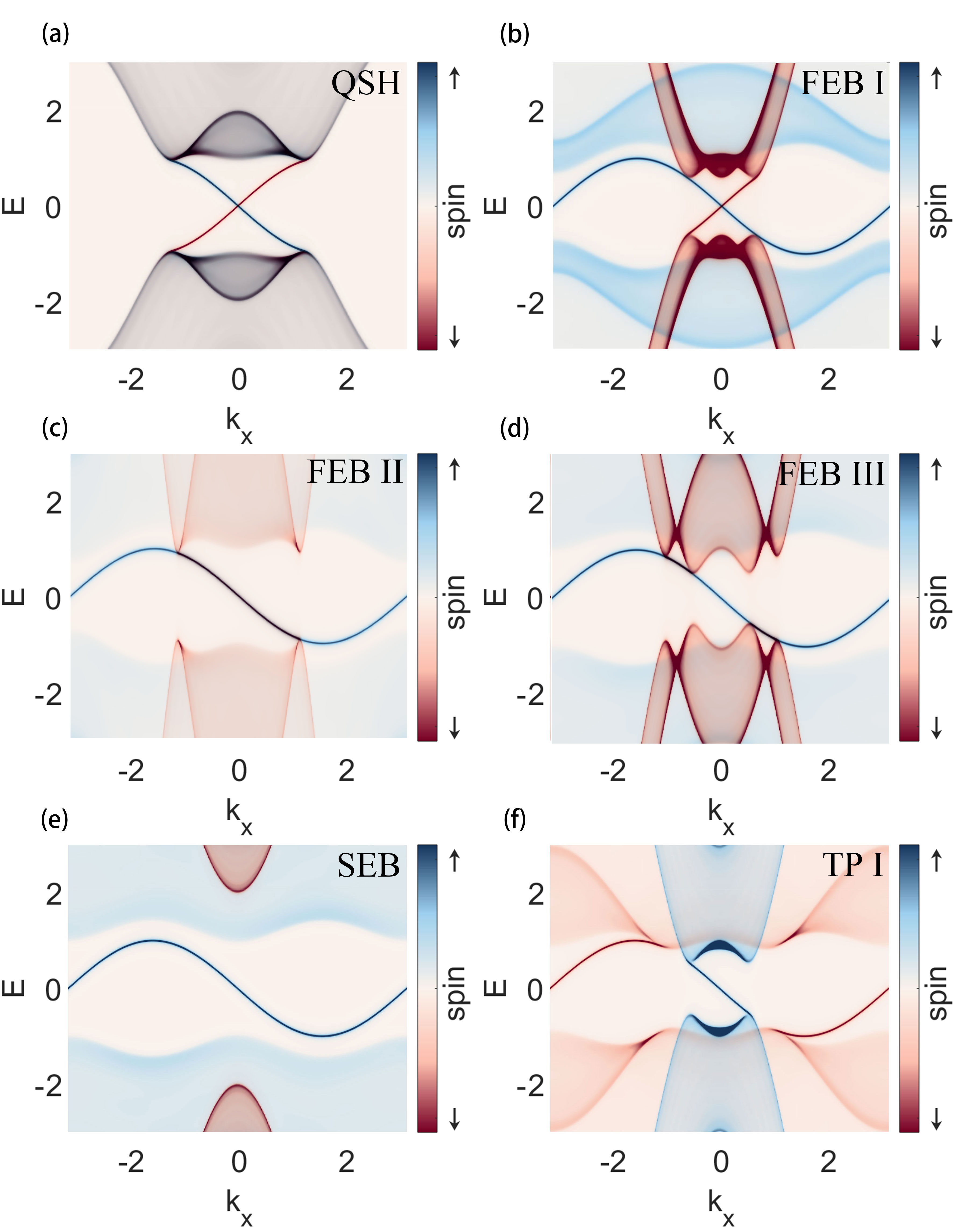}
    
    \caption{Spin-resolved density of states (DOS) along one boundary obtained by the surface Green's function method~\cite{MPLopezSancho_1985}. (a) Spin-resolved DOS for $B=1.4$ and $J=G=0$. The system is in a quantum spin Hall (QSH) phase with gapless helical edge states across the bulk gap. (b) Spin-resolved DOS for phase I in Fig.~\ref{fig:PhaseDiagram} ($B=1.4$, $J=1$, and $G=1$). (c) Spin-resolved DOS for phase II in Fig.~\ref{fig:PhaseDiagram} ($B=1.4$, $J=4$, and $G=6$). The spin-up edge bands overlap with the spin-down FEB around $k_x=0$. (d) Spin-resolved DOS for phase III in Fig.~\ref{fig:PhaseDiagram} ($B=1.4$, $J=2$ and $G=1$). The spin-up edge bands partially overlap with the spin-down FEB, while the spin-up bulk bands touch the FEB. (e) Spin-resolved  DOS for $B=1.4$, $J=1$ and $G=4$. The spin-up FEB is completely detached from the bulk bands of both spins, forming a SEB phase. (f) Spin-resolved DOS for $B=-1$, $J=3$ and $G=1$. Edge bands of different spins are located in different $k_x$ regions. This topological phase is denoted as TP I in Fig.~\ref{fig:PhaseDiagram} (d). Other parameters for panels are $A=1$ and $M=2$.}
    \label{fig:DOS}
    \end{figure}

We begin with examining the case with $-4B<M_0<0$ and $G=J=0$, where the quantum spin Hall effect emerges~\cite{Qi-Wu-Zhang06PRB}. As expected, the bulk bands keep spin degeneracy due to TRS [grey bands in Fig.~\ref{fig:DOS}(a)]. Helical states moving towards opposite directions appear on both edges along the $y$ direction, as shown by the discrete bands in Fig.~\ref{fig:DOS}(a). 
In the topological phase with winding numbers $(W_0,W_\pi)=(\pm1,0)$ as shown in Fig.~\ref{fig:PhaseDiagram}(a), the spin-up edge state keeps moving towards (backwards) $x$ direction while the magnetism lifts the spin degeneracy of bulk bands. Similarly, in the topological phase with winding numbers $(W_0,W_\pi)=(0,\pm1)$, edge bands connecting the inverted bands of spin-up electrons reach zero energy at $k_x=\pi$ and create similar edge current moving along the boundary. In contrast, the trivial phase is identified by winding numbers $(W_0,W_\pi)=(0,0)$, where no edge state is observed. 

By considering the combined topological phases of both spin blocks, we arrive at the phase diagram depicted in Fig.~\ref{fig:PhaseDiagram}(c). The solid lines represent the phase boundaries of the spin-up block, while the dashed lines indicate those of the spin-down block. Remarkably, the FEB phase, characterized by winding numbers $(W_0^{\pm},W_\pi^{\pm})=(\pm 1,\pm 1)$ in either block ("+" or "-"), emerges even though the bulk system holds a zero Chern number. This phase extends across broad ranges of the parameters, as highlighted by the color-shaded areas in Fig.~\ref{fig:PhaseDiagram}(c). When one of the spins (e.g. spin-up) enters the FEB phase while the other (spin-down) enters a topological regime with low-energy modes at $k_x$=0, the spin-down band can either form a quantum spin Hall regime with the spin-up FEB [see Fig.~\ref{fig:DOS}(b)] or overlap with the spin-up FEB [see Fig.~\ref{fig:DOS}(c)]. These two phases are labelled as phases I [characterized by ($W_0,W_\pi$)=($\pm 1,\pm 1$) for one spin and ($W_0,W_\pi$)=($\pm 1,0$) for the other] and phase II [characterized by ($W_0,W_\pi$)=($\pm 1,\pm 1$) for one spin and ($W_0,W_\pi$)=($\mp 1,0$) for the other] in Fig.~\ref{fig:PhaseDiagram}(c), respectively. The overlapping edge states can be split, for instance, by a spectrum asymmetry term $2\delta J(\cos k_y-\cos k_x)s_3\sigma_0$. Moreover, the spin-down block can transit into another topological phase, leading to another phase marked as phase III [characterized by ($W_0,W_\pi$)=($\pm 1,\pm 1$) for one spin and ($W_0,W_\pi$)=(0,0) for the other] in Fig.~\ref{fig:PhaseDiagram}(c). In this phase, the spin-down edge bands may overlap with the FEB from the spin-up block and its bulk bands may touch the FEB [see  Fig.~\ref{fig:DOS}(d)].
Thus, to obtain edge bands completely detached from the bulk bands, one block, say the spin-down block, must be driven into a trivial phase. These topological phases, where the bulk and edge bands are completely separated for the full system [see Fig.~\ref{fig:DOS}(e)], are denoted as SEB in Fig.~\ref{fig:PhaseDiagram}(c).

The appearance of the FEB phase by altermagnetism does not depend on a specific range of $M$ and $B$. We can also reach the FEB phase from the trivial topological phase at zero fields. For example, in the case with $M_0>-4B>0$, the SEB phase from the spin-down block can be obtained by introducing a Zeeman field $G$ to invert the bands at $k_x=0$ and then adding the altermagnetic field $J$ to invert the bands at $k_x=\pi$, as shown in Fig.~\ref{fig:PhaseDiagram}(d). Note that in this case, a constant Zeeman field is also crucial for the formation of the FEB. As the altermagnetic term $2J(\cos k_y-\cos k_x)$ takes different signs at $k_x=0$ and $k_x=\pi$, it alone cannot induce band inversions at both $k_x=0$ and $k_x=\pi$ to form the FEB phase. Another interesting topological phase shown in Fig.~\ref{fig:DOS}(f) appears when total winding number $(W_0,W_\pi)=(\pm1,\pm1)$ is assembled by $(W_0^+,W_\pi^+)=(\pm1,0)$ and $(W_0^-,W_\pi^-)=(0,\pm1)$. Though such a phase shares the same total winding number with the FEB phase, its edge bands do not extend to the whole Brillouin zone. The spin-up and spin-down edge bands may patch together to form a fragmented "FEB" with non-zero Chern number $C=C_++C_-=-2$. It is worth noting that when the edge bands of different spins do not overlap in $k_x$ space, the edge states on the same edge will not interact with each other even in the presence of superconductivity and in-plane magnetization [Sec.~\ref{sec:robustness}]. In the appendix, we present the explicit evolution of these states by changing the parameters $G$ and $J$ in the phase diagram.

\subsection{Edge Energy Spectrum and Wave Function \label{sec:wavefunction}}

Now, we derive the energy spectrum and the wave function of the floating edge states. 
By confining the 2D system into a semi-infinite plane (i.e., $y>0$), the wave function in the $x$ direction is a plane wave. Moreover, there is only one FEB on each edge as discussed in Sec.~\ref{subs:TopPhase}, and it suffices to consider the solution in one of the spin blocks (e.g., the spin-up block $H_+$).

To derive the floating edge states, we divide the Hamiltonian in Eq.~(\ref{eq:H}) into two parts: $H_+=H_{1D}+A\sin k_x \sigma_1$~\cite{2024arXiv240509121B}. The chiral symmetry $\mathcal{C}=\sigma_1$ is preserved in $H_{1D}$, i.e., $\{\mathcal{C},H_{1D}\}=0$. We first calculate the wave function of zero edge mode for $H_{1D}$. By showing that the zero mode is also an eigen state of the $A\sin k_x$ term, we verify that the zero mode of $H_{1D}$ captures the wave function of the full Hamiltonian $H_+$ with energy $E=\pm A\sin k_x$.
The zero modes of $H_{1D}$ share the same wave functions with the chiral symmetry operator, which significantly simplifies our calculation.

 \begin{figure}[t]
    \includegraphics[width=1.02\linewidth]{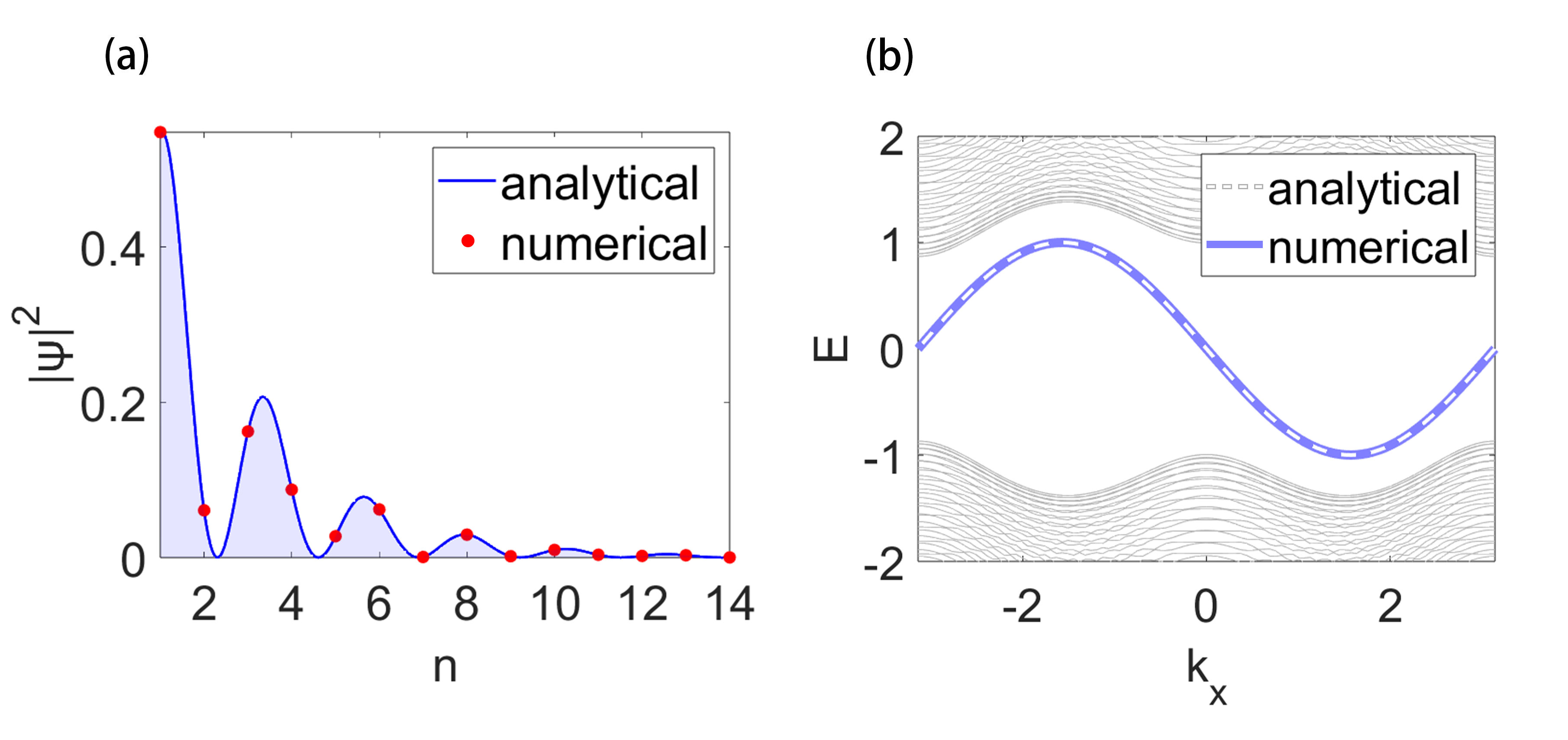}
    
    \caption{(a) Analytical (blue solid line) and numerical (red dots) results of an edge-state wave function with momentum $k_x=\pi/4$ in the FEB. (b) Analytical (white dashed line) and numerical (blue solid line) results of the energy spectrum of the FEB. The grey region indicates the bulk continuum obtained numerically. Other parameters are $M=2,B=1.4,A=1,J=1$ and $G=2$. 
    }
    \label{fig:Edge-Band}
    \end{figure}

To proceed, we rewrite the wave function of $H_{1D}$ in the basis $\Psi=(c_{1,\uparrow,a,k_x},c_{1,\uparrow,b,k_x},\dots,c_{n,\uparrow,a,k_x},c_{n,\uparrow,b,k_x},\dots)^T$. The corresponding Hamiltonian then reads
\begin{widetext}
\begin{equation}
    H_{1D}=\begin{pmatrix}
        m(k_x)&0&B+J&-{A}/{2}&0&0&\dots\\
        0&-m(k_x)&{A}/{2}&-B-J&0&0&\dots\\
        B+J&{A}/{2}&m(k_x)&0&B+J&-{A}/{2}&\dots\\
        -{A}/{2}&-B-J&0&-m(k_x)&{A}/{2}&-B-J&\dots\\
        0&0&B+J&{A}/{2}&m(k_x)&0&\dots\\
        0&0&-{A}/{2}&-B-J&0&-m(k_x)&\dots\\
        \vdots&\vdots&\vdots&\vdots&\vdots&\vdots&\ddots
    \end{pmatrix},
\end{equation}
where $m(k_x)=M_0+G+2(B-J)\cos k_x$. Inserting $\ket{\Psi}=(\psi_{1,a},\psi_{1,b},\dots,\psi_{n,a},\psi_{n,b},\dots)^T$ into the eigen equation $H_{1D}\ket{\Psi}=0$ for the zero mode (suppose it exists), we arrives at the following iterative equations:
\begin{equation}
\begin{cases}
    m(k_x)\psi_{1,a}+(B+J)\psi_{2,a}-(A/2)\psi_{2,b}&=0,\\
    -m(k_x)\psi_{1,b}+(A/2)\psi_{2,a}-(B+J)\psi_{2,b}&=0,\\
    (B+J)\psi_{1,a}+(A/2)\psi_{1,b}+m(k_x)\psi_{2,a}+(B+J)\psi_{3,a}-(A/2)\psi_{3,b}&=0,\\
    -(A/2)\psi_{1,a}-(B+J)\psi_{1,b}-m(k_x)\psi_{2,b}+(A/2)\psi_{3,a}-(B+J)\psi_{3,b}&=0,\\
    \; \;\;\;\;\;\; \qquad \dots\\
    (B+J)\psi_{n-1,a}+(A/2)\psi_{n-1,b}+m(k_x)\psi_{n,a}+(B+J)\psi_{n+1,a}-(A/2)\psi_{n+1,b}&=0,\\
    -(A/2)\psi_{n-1,a}-(B+J)\psi_{n-1,b}-m(k_x)\psi_{n,b}+(A/2)\psi_{n+1,a}-(B+J)\psi_{n+1,b}&=0, \\ 
    \; \;\;\;\;\;\; \qquad \dots
    \label{eq:iter}
\end{cases}
\end{equation}
\end{widetext}

The eigen wave functions of the chiral symmetry operator are given by
\begin{equation}
    \ket{\nu =\pm 1}=\frac{1}{\sqrt{2}}\begin{pmatrix}
        1\\\pm 1
    \end{pmatrix}=\frac{1}{\sqrt{2}}\begin{pmatrix}
        1\\ \nu
    \end{pmatrix}.
\end{equation}
It is easy to check that $(\psi_{n,a},\psi_{n,b})^T\propto \ket{\nu=\pm 1}$ satisfy Eq.~(\ref{eq:iter}). Thus, we assume the ansatz of the wave functions in the form $(\psi_{n,a},\psi_{n,b})^T=\frac{c}{\sqrt{2}}\xi^n (1,\nu)^T$. The iterative equations for $n\geq2$ are simplified to
\begin{equation}
    \Big(B+J-\nu\frac{A}{2}\Big)\xi^2+m(k_x)\xi+\Big(B+J+\nu\frac{A}{2}\Big)=0.
\end{equation}
Solving this equation yields
\begin{equation}
    \xi_{\pm}=\frac{-m(k_x)\pm\sqrt{m(k_x)^2-4(B+J)^2+A^2}}{2(B+J)-\nu A}.
\end{equation}
The wave function can be written generally as $(\psi_{n,a},\psi_{n,b})^T=(c_+\xi_+^n+c_-\xi_-^n)\ket{\nu=\pm 1}$. 
 We determine the coefficients $c_\pm$ by applying the boundary conditions. Plugging this wave function into the boundary equation, i.e. the first equation in Eq.~(\ref{eq:iter}), we obtain
\begin{equation}
    (c_+\xi_++c_-\xi_-)m(k_x)+(c_+\xi_+^2+c_-\xi_-^2)\Big(B+J-\nu \frac{A}{2}\Big)=0,
\end{equation}
which gives $c_+=-c_-$. Thus, the wave function of a zero mode of $H_{1D}$ is given by
\begin{align}
 (\psi_{n,a},\psi_{n,b}) = N (\xi_+^n-\xi_-^n) (1,\nu),
    \label{eq:WF}
\end{align}
where $N=1/\sqrt{2\sum_{n\geq1}|\xi_+^n-\xi_-^n|^2}$ is the normalization factor. 

Equation~\eqref{eq:WF} shows that the wave function is localized at the boundary only when both $|\xi_\pm|<1$. For $m(k_x)^2-4(B+J)^2+A^2<0$, i.e. $\xi_+=\xi_-^*$, this requires 
\begin{equation}
    \nu A[2(B+J)+A]<0.
\end{equation}
In this case, the wave function can be simplified as
\begin{align}
 (\psi_{n,a},\psi_{n,b}) = N \xi^n \sin (n\varphi) (1,\nu),
    \label{eq:WF2}
\end{align}
where $N=1/\sqrt{2\sum_{n\geq1}\xi^{2n}|\sin(n\varphi)|^2}$, and we define
\begin{equation}
\begin{aligned}
    \xi= & \sqrt{[{2(B+J)+\nu A}]/[{2(B+J)-\nu A}]}, \\
    \varphi= & \arctan\big({\sqrt{4(B+J)^2-A^2-m(k_x)^2}}/{m(k_x)}\big).
\end{aligned}
\end{equation}
To be more specific, in the FEB topological region, e.g., with $B>0$ and $|G+M_0|<4 \min(B,J)$, we find that for all $k_x$, $m(k_x)^2-4(B+J)^2+A^2<0$ holds, while the norm of $\xi_\pm$ satisfies $|\xi_{\pm}|^2=(2B+2J+\nu A)/(2B+2J-\nu A)<1$ for $\nu=-\text{sgn}(AJ)$. The wave function takes the form of a decaying sine wave. Interestingly, we find that while the periodicity $2\pi/\varphi$ of the oscillations depends strongly on the momentum $k_x$, the decay behavior determined by $\xi$ is constant in $k_x$.  Moreover, the decay length (or $\xi$) appears to be independent of the bulk gap $M$ at $\Gamma$, in stark contrast to the edge states in quantum spin Hall insulators, where the localization length is directly determined by the bulk gap.  This constant decay behavior may constitute a unique characteristic of the floating edge states throughout the Brillouin zone.  

Note that importantly the wave function in Eq. (\ref{eq:WF2}) is also the eigenstate of $A\sin k_x \sigma_1$ with eigenvalue $\nu A\sin k_x$. Therefore, the zero modes of $H_{1D}$ are indeed the edge states of the full Hamiltonian. Taking into account the $A\sin k_x \sigma_1$ part, the energy spectrum of the floating edge states is obtained as 
\begin{equation}
  E_{\text{FEB}}= \bra{\Psi} A \sin k_x \sigma_1 \ket{\Psi}= \nu A\sin k_x.
\end{equation}
The agreement of our analytical with numerical results for the wave function and energy spectrum are shown in Fig.~\ref{fig:Edge-Band}. Notably, the inclusion of the $A\sin k_x\sigma_x$ part only changes the energy of the FEB but does not change the details of the wave function. Thus, the wave function of the FEB remains a decaying sine wave along the $y$ direction, with spin-up currents along each edge.

\section{Stability of the Topological Phases \label{sec:robustness}}

So far, we have demonstrated the emergence of FEB in a ribbon aligned along the $x$ direction, assuming the spectral symmetry and decoupled spin species. In the following, we will show that the FEB phase is more generic and persists even when these constraints are lifted. Specifically, we will prove that the FEB remains robust against various perturbations, including rotations of the ribbon orientation, energy spectrum asymmetries, and spin coupling, respectively. 

\begin{figure}
\includegraphics[width=1\linewidth]{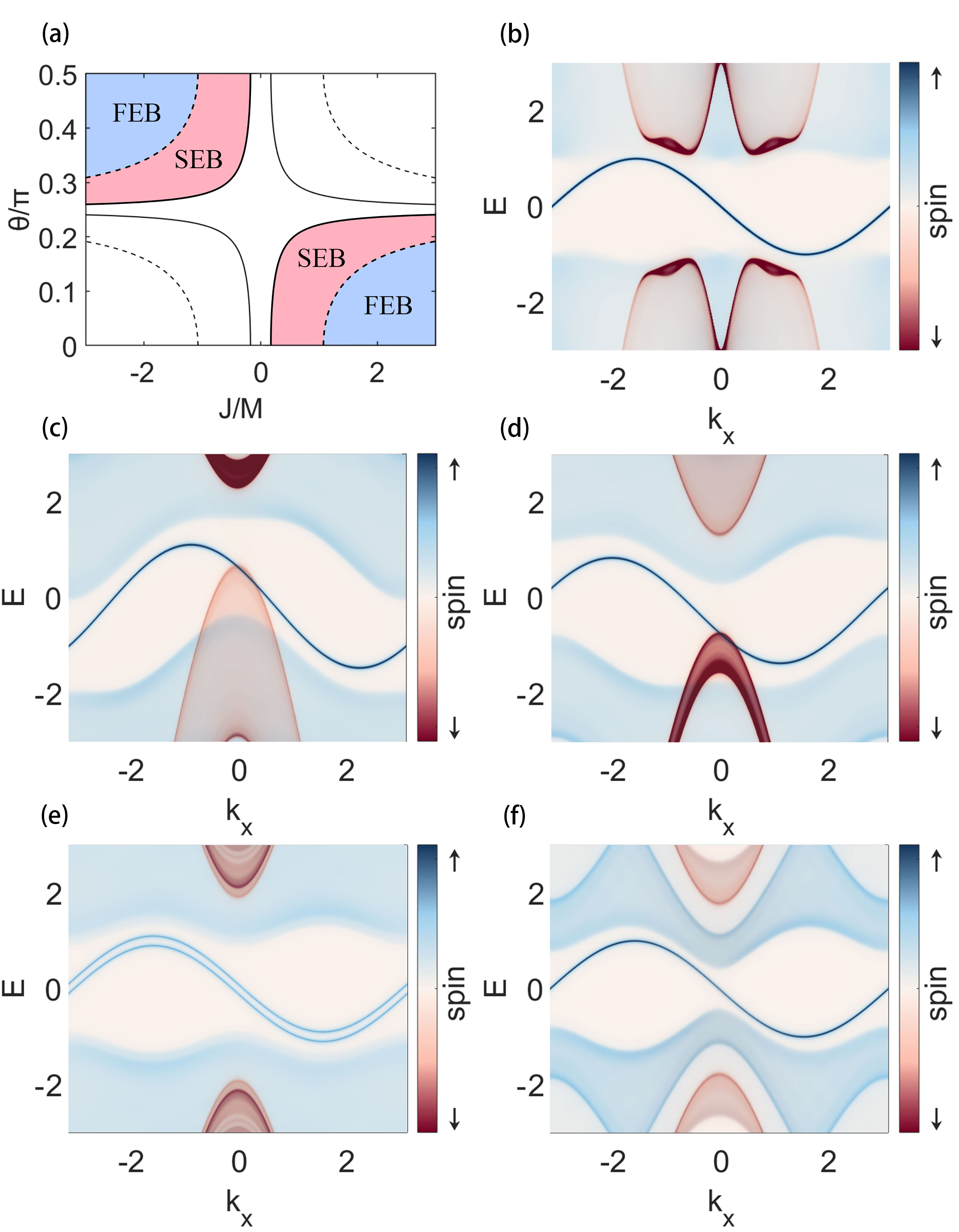}
    
\caption{(a) Phase diagram against the strength $J$ and orientation $\theta$ of the altermagnetism for $G=5$. The topological phases vanish when the altermagnetism vanishes effectively at $\theta=\pi/4$. 
(b) Spin-resolved DOS for $J=-4$, $G=5$ and $\theta=0.3\pi$. The FEB is identical to that for $J=-4\cos 2\theta=1.24$ with distorted bulk bands. (c) Spin-resolved DOS for $J=1$ and in the presence of a spectrum asymmetry term $\epsilon(\bf k)$ (i.e., $G=3$, $\mu=1.7$ and $t=-0.5$). The $\epsilon(\bf k)$ term acts as a distortion and preserves the FEB phase. (d) Spin-resolved DOS for $J=1$, $G=3$, $\delta J/J=0.2$ and $\delta G/G=-0.1$. The FEB phase persists even in the presence of energy asymmetries. Other parameters are $M=2,B=1.4$ and $A=1$. (e) Spin-resolved DOS for $J=1$, $G=4$, $\mu=0.1$ and an $s$-wave pairing potential with strength $\Delta=0.1$. (f) Spin-resolved DOS for $J=0.5$, $G=3$, and additional Zeeman fields $G_x=1$, $G_y=2$ in the $x$ and $y$ directions, respectively.
}
\label{fig:Stablility}
\end{figure}

\subsection{Ribbon with a Rotational Angle}

We first study the influence of a rotation of the ribbon (edge) orientation. In the absence of altermagnetism, the BHZ model has an effective isotropic property at low energies. This may allow us to use an effective rotated Hamiltonian to estimate the result instead of considering the exact cutting of complex edges. To be more specific, we derive the $k\cdot p$ Hamiltonian of Eq. (\ref{eq:H}) as
\begin{equation}
\begin{aligned}
    H_{k\cdot p}({\bf k})= &\; (M-Bk^2)\sigma_3+A(k_xs_3\sigma_1+k_y\sigma_2)\\
    & +J(k_x^2-k_y^2)s_3\sigma_3+G s_3\sigma_3.
\end{aligned}
\end{equation}
After the rotation of $\bf{k}$, i.e., $(k_x,k_y)^T\to R(k_x,k_y)^T=e^{-i\theta\sigma_2}(k_x,k_y)^T$, the altermagnetic term becomes
\begin{equation}
\begin{aligned}
    & J\begin{pmatrix}
        k_x&k_y
    \end{pmatrix}R\begin{pmatrix}
        1&0\\0&-1
    \end{pmatrix}R^{-1}\begin{pmatrix}
        k_x\\k_y
    \end{pmatrix}\\
    = & \; J[(k_x^2-k_y^2)\cos (2\theta) + 2k_xk_y\sin (2\theta)].
\end{aligned}
\end{equation}
The $B$ and $A$ terms are isotropic and therefore remain unchanged under the rotation.
Then, we can revert to the lattice model as follows:
\begin{equation}
\begin{aligned}
  H(\theta)= & M({\bf k})\sigma_3+A(\sin k_x s_3\sigma_1 + \sin k_y  \sigma_2)\\
  & +2J[(\cos k_y -\cos k_x)\cos 2\theta\\
  & +\sin k_x\sin k_y \sin 2\theta]s_3\sigma_3 + G s_3 \sigma_3,
  \label{eq:Hr}
\end{aligned}
\end{equation}
where $M({\bf k})$ is defined as in Eq. (\ref{eq:H}).

Following the same procedure as in Sec.~\ref{sec:wavefunction}, we find that the emergent chiral symmetry is still preserved in Eq.~\eqref{eq:Hr} at $k_x=0$ and $k_x=\pi$. At both points, $H(\theta)$ retains the same form as $H(\theta=0)$ but with $J$ replaced by $J\cos (2\theta)$. The typical phase diagram as a function of $J$ and $\theta$ is shown in Fig.~\ref{fig:Stablility}(a). We observe that the FEB phase remains in broad parameter regimes as indicated by the colored areas. Moreover, it reflects the weak topological nature of the system, where the FEB phase disappears at $\theta=\pi/4$, corresponding to an effective vanishing of altermagnetism along the edge. By rotating the ribbon (edge) orientation, the region of the SEB phase appears to expand. However, it is important to note that while the $\sin 2\theta$ term will not affect the phase boundary, it distorts the bulk bands, as illustrated in Fig.~\ref{fig:Stablility}(b).

\subsection{Spectrum Asymmetry and Spin Coupling}
Next, we consider the case with spectrum asymmetries. To this end, we extend the Hamiltonian in Eq.~(\ref{eq:H}) to
\begin{equation}
\begin{aligned}
  H_{ext}({\bf k})= &\; \epsilon({\bf k}) +M({\bf k})\sigma_3+A(\sin k_x s_3\sigma_1 + \sin k_y \sigma_2)\\
  & +2J(\cos k_y -\cos k_x)s_3\sigma_3 + G s_3 \sigma_3 \\
  & +2\delta J(\cos k_y -\cos k_x)s_3 + \delta G s_3.
\end{aligned}
\end{equation}
Here, $\epsilon({\bf k})=\mu+4t-2t(\cos k_x + \cos k_y)$ accounts for the electron-hole asymmetry that usually occurs in the BHZ model. The terms associated with $\delta J$ and $\delta G$ capture the effects of the $\sigma_0$ coupling, arising due to the asymmetry between the magnetic response of the two orbitals. 

In the orbital space, these additional terms act as constants, which only modify the eigenenergies without altering the overall phase diagram. This can be verified through numerical calculations.  Figures~\ref{fig:Stablility}(c) and \ref{fig:Stablility}(d) present the distorted FEB under the presence of the $\epsilon$ term and the $\delta J$ and $\delta G$ terms, respectively. Although the spin-up edge bands may partially overlap with the spin-down bulk bands, they can still be distinguished through spin identification.
Furthermore, since there is only one edge state localized at each edge of the ribbon, the overlap between the two edge states at opposite edges is proportional to $\sim |\xi|^{L_y}$. Thus, the coupling between the two spin species (e.g., by in-plane magnetism or singlet superconductivity) will not significantly distort the bands as long as $L_y\gg |\xi|$, a condition that is easily met. The numerical results are presented in Figs.~\ref{fig:Stablility}(e) and \ref{fig:Stablility}(f) for the cases of additional local superconductiviting pairing potential and in-plane Zeeman field, respectively. Although the distortion of bulk bands will affect the gap closing point, we still have a robust FEB phase in a wide parameter range with perturbative spin coupling.

\section{Summary \label{sec:summary}}

We have shown that the FEB phase with complete detachment of bulk and surface bands can be realized in the BHZ model under $d$-wave altermagnetic and Zeeman fields for broad parameter regimes. We have obtained the exact topological phase diagrams of the system and analytical results of the energy spectrum and wave functions of the floating edge states. Furthermore, we have verified the robustness of the FEB phase under various perturbations such as geometric rotations, spectrum asymmetries, and spin coupling.

\begin{acknowledgments}

S.B.Z. acknowledges the support of the start-up fund at HFNL, the Innovation Program for Quantum Science and Technology (Grant No. 2021ZD0302800), and Anhui Initiative in Quantum Information Technologies (Grant No. AHY170000).
 
\end{acknowledgments}


%
    
\appendix

\onecolumngrid

\section{Evolution of Topological Phases}\label{Appendix:Evolution}

We present the evolution of the topological phases in detail. Transitions between distinct topological phases are marked by gap closings at either $k_x=0$ or $k_x=\pi$. To illustrate this, we examine the cases of $-4B<M_0<0$ and $M_0>-4B>0$, respectively, where we define $M_0=M-4B$ for brevity.\\

\begin{figure}[b]
    \centering
    \includegraphics[width=.9\linewidth]{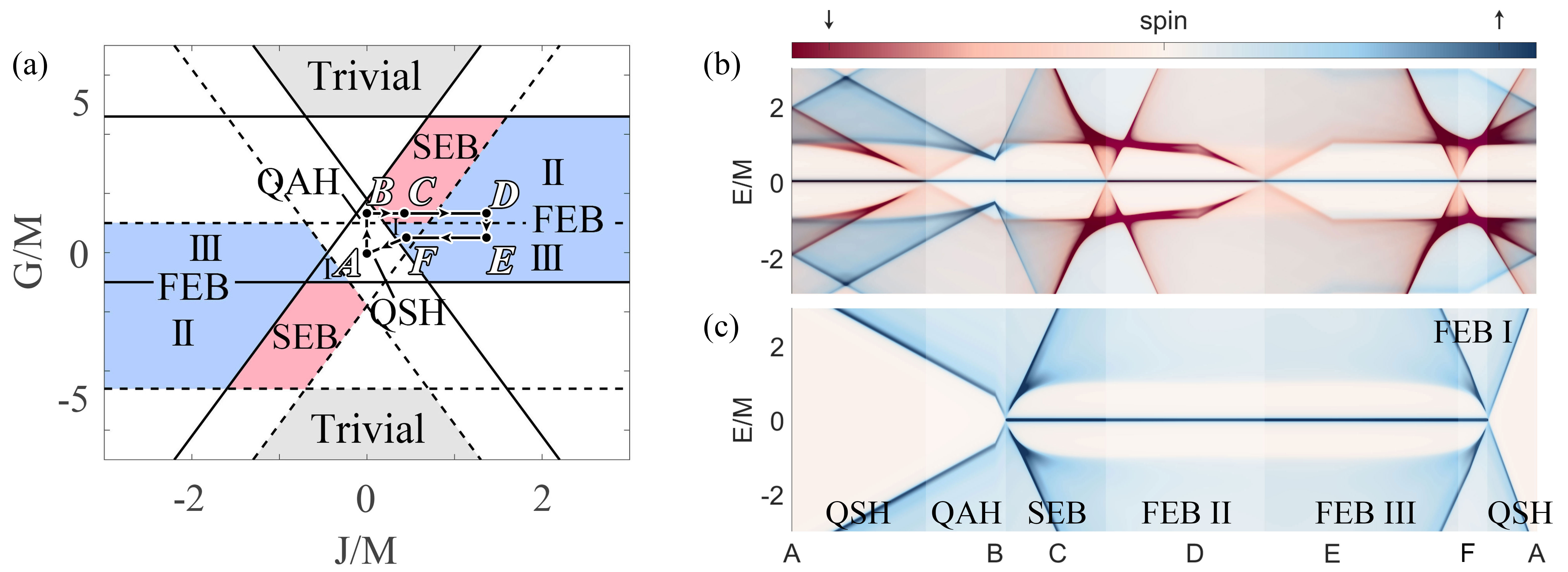}
    
    \caption{(a) The evolution path $A\to B\to C\to D\to E\to F\to A$ that traces the transitions between topological phases in the phase diagram. (b,c) The band structure evolution along the loop path marked in (a). Phase transitions occur when the bulk gap closes at (b) $k_x=0$ or (c) $k_x=\pi$. Other parameters are $M=2$, $B=1.4$, $A=1$.}
    \label{fig:Loop1}
\end{figure}

  \begin{figure}[t]
    \centering
    \includegraphics[width=.9\linewidth]{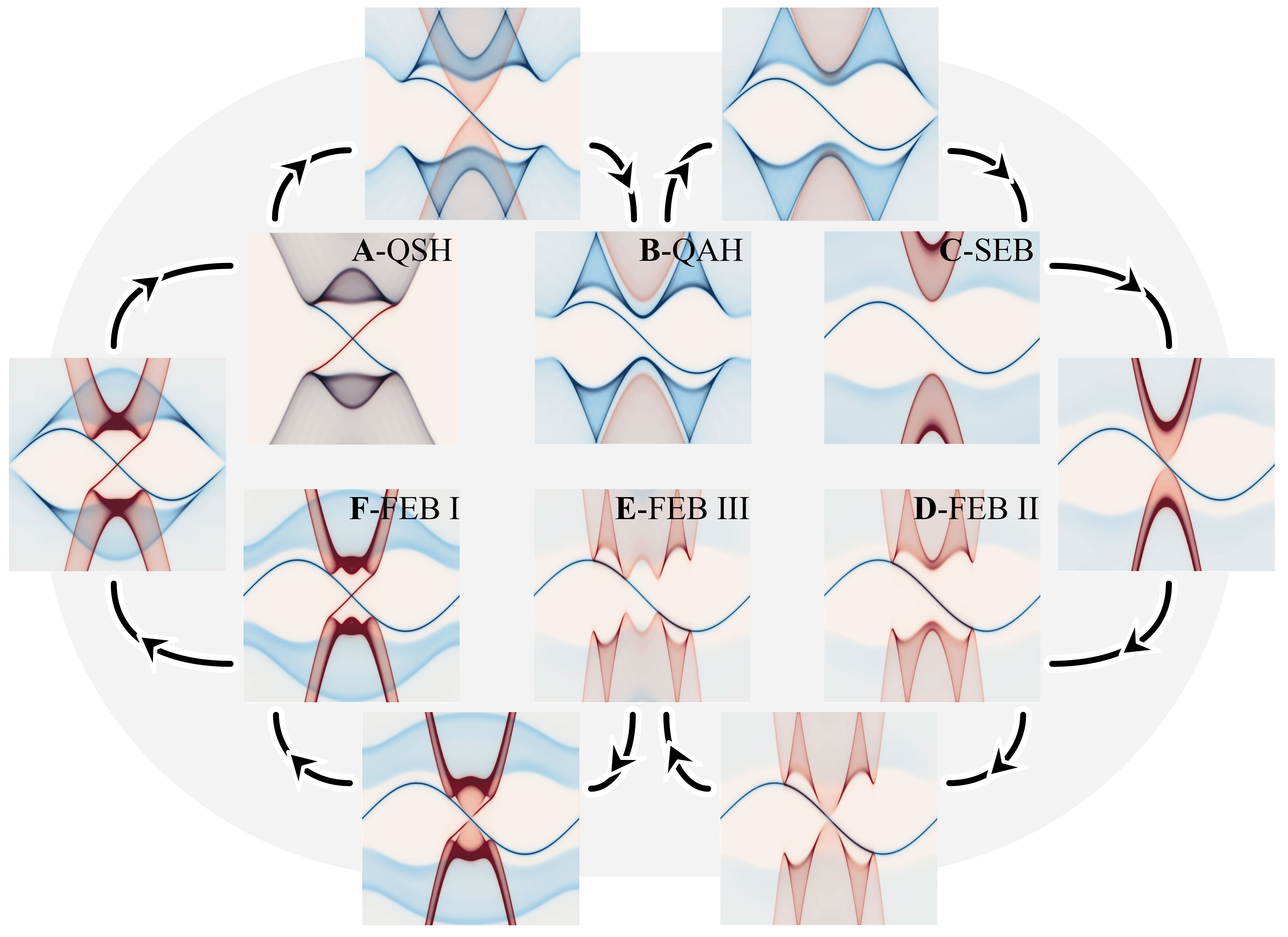}
    
    \caption{Spin-resolved DOS for (A) QSH phase with $(J,G)=(0,0)$, (B) QAH phase with $(J,G)=(0,3)$, (C) SEB phase with $(J,G)=(1,3)$, (D) FEB II phase with $(J,G)=(3,3)$, (E) FEB III phase with $(J,G)=(3,1)$, and (F) FEB I phase with $(J,G)=(1,1)$. The other panels display the gap-closing points at transitions between these phases. Other parameters are $M=2$, $B=1.4$, and $A=1$.}
    \label{fig:Evolution+}
    \end{figure}

\textbf{The case of $-4B<M_0<0$:}
The phase evolution along the loop path $A\to B\to C\to D\to E\to F\to A$, marked in Fig.~\ref{fig:Loop1}(a), is illustrated in Figs.~\ref{fig:Loop1}(b) and \ref{fig:Loop1}(c). The phase transitions occur when the bulk gap closes at $k_x=0$ [see Fig.~\ref{fig:Loop1}(b)] or $k_x=\pi$ [see Fig.~\ref{fig:Loop1}(c)].    
To be specific, we start with the quantum spin Hall (QSH) phase (point A) at $J=G=0$ in Fig.~\ref{fig:Evolution+}A and introduce the Zeeman field $G$. The bulk gap closes when the path crosses the phase boundary at $G=M$, removing the edge state for one spin and entering the quantum anomalous Hall (QAH) phase (point B). Next, we turn on the altermagnetism $J$, which closes the gap at the phase boundary $G+4J=-M_0$, and enter the SEB phase (point C), creating an additional low-energy spin-up edge state at $k_x=\pi$. From the SEB phase, as we further increase the altermagnetic strength, the gap closes when the path crosses the phase boundary at $G-4J=M_0$, entering the FEB II phase (point D) with spin-down edge modes overlapping with the spin-up FEB. Subsequently, we reduce the strength of the Zeeman field. The bulk gap closes when the path crosses the phase boundary at $G=M$, removing the spin-down edge modes around $k_x=0$ and transitioning the system into the FEB III phase (point E). Reducing the altermagnetic field further results in a gap closing a $G-4J=M_0$, creating spin-down edge modes at $k_x=0$ and entering the FEB I phase (point F). Finally from point F in the FEB I phase, we turn off both altermagnetic and Zeeman fields. The bulk gap closes when the path crosses the phase boundary at $G+4J=-M_0$, removing the spin-up edge modes at $k_x=\pi$ and returning to the QSH phase (point A).\\

  \begin{figure}[H]
    \centering
    \includegraphics[width=.9\linewidth]{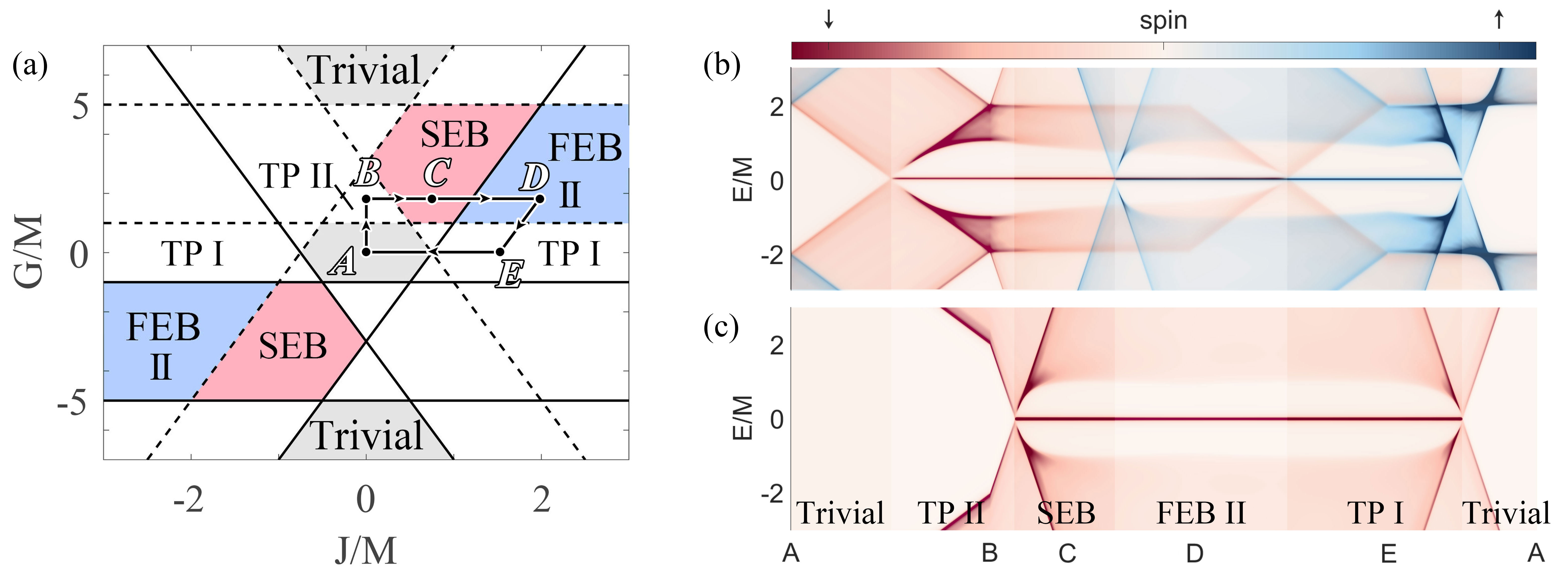}
    
    \caption{(a) The evolution path $A\to B\to C\to D\to E\to A$ that traces the transitions between topological phases in the phase diagram. (b,c) The band structure evolution along the loop path marked in (a). Phase transitions occur when the bulk gap closes at (b) $k_x=0$ or (c) $k_x=\pi$. Other parameters are $M=2$, $B=-1$, $A=1$.}
    \label{fig:Loop2}
    \end{figure}

  \begin{figure}[H]
    \centering
    \includegraphics[width=.9\linewidth]{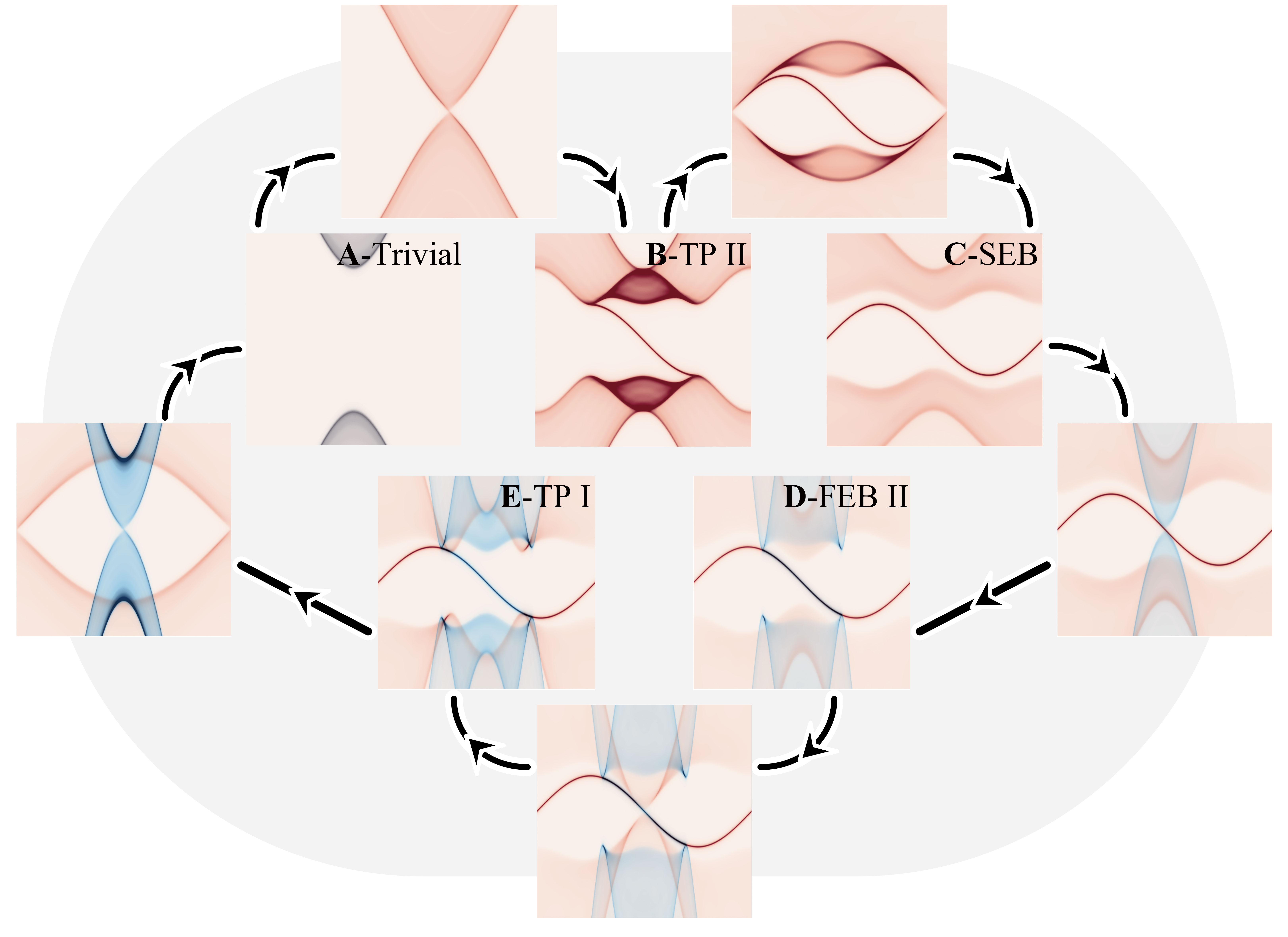}
    
    \caption{Spin-resolved DOS for (A) trivial phase with $(J,G)=(0,0)$, (B) TP II phase with $(J,G)=(0,4)$, (C) SEB phase with $(J,G)=(1.5,4)$, (D) FEB II phase with $(J,G)=(4,4)$, and (E) TP I phase with $(J,G)=(3,0)$. The other panels display the gap-closing points at transitions between these phases. Other parameters are $M=2$, $B=-1$, and $A=1$.}
    \label{fig:Evolution-}
    \end{figure}

\textbf{The case of $M_0>-4B>0$:} 
The phase evolution along the loop path $A\to B\to C\to D\to E\to A$, marked in Fig.~\ref{fig:Loop2}(a), is illustrated in Figs.~\ref{fig:Loop2}(b) and \ref{fig:Loop2}(c) The phase transitions occur when the bulk gap closes at  $k_x=0$ [see Fig.~\ref{fig:Loop2}(b)] or $k_x=\pi$ [see Fig.~\ref{fig:Loop2}(c)]. 
To be specific, we start with the trivial phase (point A) at $J=G=0$ in Fig.~\ref{fig:Evolution-}A and introduce the Zeeman field $G$. The bulk gap closes when the path crosses the phase boundary at $G=M$, creating spin-down edge modes at $k_x=0$ and entering the topological (TP) II phase (point B). Next, we turn on the altermagnetism to close the gap when the path crosses the phase boundary at $G+4J=M_0$, further creating another spin-down edge mode at $k_x=\pi$ and transitioning the system into the SEB phase (point C). From the SEB phase, we then increase the altermagnetic strength, the gap closes when the path crosses the phase boundary at $G-4J=-M_0$, transitioning the system into the FEB II phase (point D) by creating spin-up edge modes at $k_x=0$. Subsequently, we reduce the strengths of both magnetic fields. The gap closes when the path crosses the phase boundary at $G=M$,  removing spin-down edge modes around $k_x=0$ and transitioning the system into the TP I phase (point E). Finally, we turn off the altermagnetic field and close the gap when the path crosses the phase boundary at $4J=M_0$, removing the edge modes and returning the system to the trivial phase (point A).

\end{document}